\documentclass[twocolumn,showpacs,preprintnumbers,amsmath,amssymbo]{revtex4}
\usepackage{graphicx}
\usepackage{dcolumn}
\usepackage{bm}
\usepackage{amsmath}
\usepackage{textcomp}
\usepackage{hyperref}
\usepackage{amssymb}
\usepackage{amsfonts}
\usepackage{mathrsfs}
\usepackage[usenames,dvipsnames]{color}

\begin{document}
   \title{Gravitomagnetism in Quantum Mechanics}
    \author{Ronald J. Adler}
     \affiliation{Gravity Probe B, Hansen Laboratory for Experimental Physics, Stanford University, Stanford California
94309}
    \author{Pisin Chen}%
     \affiliation{1. Leung Center for Cosmology and Particle Astrophysics \& 
                   Department of Physics and Graduate Institute of Astrophysics, National Taiwan University, Taipei, Taiwan 10617\\
                  2. Kavli Institute for Particle Astrophysics and Cosmology, SLAC National Accelerator Laboratory, Menlo Park, CA 94025}


\begin{abstract}
We give a systematic treatment of the quantum mechanics of a spin zero particle in a combined electromagnetic field and a weak gravitational field, which is produced by a slow moving matter source. The analysis is based on the Klein-Gordon equation expressed in generally covariant form and coupled minimally to the electromagnetic field. The Klein-Gordon equation is recast into Schroedinger equation form (SEF), which we then analyze in the non-relativistic limit. We include a discussion of some rather general observable physical effects implied by the SEF, concentrating on gravitomagnetism. Of particular interest is the interaction of the orbital angular momentum of the particle with the gravitomagnetic field.
\end{abstract}

\maketitle

\renewcommand{\thesection}{\arabic{section}}
\renewcommand{\theequation}{\thesection.\arabic{equation}}

\section{\label{sec:level1}Introduction}

Over the course of centuries there has been enormous effort devoted to theories and experiments for classical systems in external gravitational fields \cite{1.1}\cite{1.2}\cite{1.3}\cite{1.4}. Two recent experiments purport to give evidence for a gravitomagnetic field, which is the gravitational analog of a magnetic field due to the motion of source matter: the LAGEOS experiments detect gravitomagnetic effects on the orbital motion of two earth satellites, and the Gravity Probe B experiment detects gravitomagnetic effects on the precession of four gyroscopes in earth orbit \cite{1.5}\cite{1.6}.

Much attention has also been given to the theory of quantum fields in classical background gravitational fields, in particular regarding Hawking radiation by black holes \cite{1.7}. Notable work has been done on quantum systems in classical background gravitational fields, for example on neutrons in the earth's field \cite{1.75}\cite{1.76}, and more recently for atomic beam interferometry; this last work is largely based on semi-classical calculations of phase shifts along the classical atomic trajectories \cite{1.8}. Much less work has been done on gravitomagnetic effects on fully quantum mechanical systems. Experiments to detect such effects should clearly be expected to be quite difficult, but would be of fundamental interest \cite{1.85}\cite{1.86}.

In this work we give a systematic treatment of a scalar or spin zero quantum particle in a combined electromagnetic and weak gravitational field. We describe the particle with the generally covariant Klein-Gordon equation, minimally coupled to the electromagnetic field in standard fashion \cite{1.9}. The weak gravitational field is naturally treated according to linearized general relativity, and we also limit ourselves to slowly moving sources. In this case the gravitational field equations are quite analogous to those of classical electromagnetism \cite{1.10}; we refer to this as the gravitoelectromagnetic or GEM limit.

We proceed by first casting the Klein-Gordon equation in an exact Schroedinger equation form (SEF), which curiously does not seem to appear in the literature \cite{1.11}. The SEF of course lends itself well to considerations of the non-relativistic limit for the quantum particle, which we then study to leading order in the relevant energies \cite{1.12}. The reader not interested in the theoretical development
may skip directly to the results in section 5, in particular Eq. (5.1).

Our main focus is on gravitomagnetic effects on a quantum system, so we briefly consider a number of possibly observable physical effects, in particular using a fictitious spin zero electron bound to an atom in a gravitomagnetic field. (That is we ignore spin.) It is conceivable that some novel interactions we derive, involving the product of the EM
vector potential and the gravitomagnetic field, might be of interest. In considering such physical effects we only present parametrizations and rough numerical estimates since it is beyond the scope of the present work to analyze specific experiments; we leave such considerations as a challenge to experimentalists.

In a subsequent paper we will consider spin 1/2 particles and compare the results to those of the present work, especially those related to gravitomagnetic interactions with spin angular momentum.

It should be emphasized that here we treat gravity entirely classically, so our work does not relate to quantum gravity or quantum spacetime \cite{1.1}\cite{1.13}.

\section{\label{sec:level1}Gravity in the gravitoelectromagnetic (GEM) limit}
In this section we review very briefly linearized general relativity for slowly moving matter sources such as the earth, paying special attention to the gravitomagnetic part of the field \cite{1.2}\cite{1.10}. Note that it is the gravitational source matter that we assume is moving slowly, not the quantum objects in the field. We use $\hbar=c=1$, but retain dimensions for $G$.

For a weak gravitational field the metric may be written in Lorentz coordinates as
\begin{align}
\mbox{g} _{\mu\nu}=\eta_{\mu\nu}+h_{\mu\nu},
\end{align}
where $\eta_{\mu\nu}$ is the Lorentz metric and $h_{\mu\nu}$ is a small perturbation.
For ordinary matter such as that of the earth the energy momentum tensor is well approximated by the matter or ``dust'' tensor \cite{2.1},
\begin{align}
T_{\mu\nu}=\rho u_{\mu}u_{\nu}.
\end{align}
Here $\rho$ is the scalar density of the source matter and $u_\mu$ is its 4-velocity. To first order in $h_{\mu\nu}$ the field equations are \cite{1.10}
\begin{align}
\partial^2(h_{\mu\nu}-\frac{1}{2}\eta_{\mu\nu}h)=-16\pi G\rho u_{\mu}u_{\nu}.
\end{align}
Here $\partial^2=\partial_t^2-\nabla^2$ is the d'Alembertian operator and we have used coordinate freedom to impose the Lorentz condition
\begin{align}
(h_{\mu\nu}-\frac{1}{2}\eta_{\mu\nu}h)^{|\mu}=0.
\end{align}
The single slash denotes an ordinary derivative.

For slowly moving source matter we may ignore second and higher order terms in the source velocity. Then the linearized field Eq.s (2.3) imply a simple form for $h_{\mu\nu}$
\begin{align}
h_{\mu\nu}=
\begin{pmatrix}
  2\phi & h^1   & h^2   & h^3\\
  h^1   & 2\phi & 0     & 0  \\
  h^2   & 0     & 2\phi & 0  \\
  h^3   & 0     & 0     & 2\phi
\end{pmatrix},\mbox{ }
h^k\equiv h_{0k},
\phi\equiv h_{00}/2,
\end{align}
where $\phi$ is the Newtonian or gravitoelectric potential, and $h^k$ is the gravitomagnetic potential. The fields $\phi$ and $h_k$ obey
\begin{subequations}
\begin{align}
-\partial^2\phi=4\pi G\rho,\mbox{ } -\partial^2h^j=-16\pi G\rho \mbox{v}^j,\\
4\dot{\phi}-\nabla \cdot \vec{h}=0,\mbox{ }\dot{\vec{h}}=0.
\end{align}
\end{subequations}
Equations (2.6a) are the field equations, while Eq.s (2.6b) follow from the Lorentz conditions. For low velocity sources we expect $\dot{\phi}\approx\phi(v/r)$ and $\ddot{\phi}\approx\phi(v^2/r^2)$ where
$r$ is the characteristic distance to the source, so we will ignore $\ddot{\phi}$ in Eq.s (2.6) and everywhere else henceforth, but we will retain $\dot{\phi}$. Similarly we will ignore both $\dot{h}^j$ and $\ddot{h}^j$. Then Eq.s (2.6) become
\begin{align}
\nabla^2\phi=4\pi G\rho,\mbox{ } \nabla^2h^j=-16\pi G\rho \mbox{v}^j,\mbox{ }4\dot{\phi}-\nabla \cdot \vec{h}=0.
\end{align}
We refer to Eq.s (2.7), with slowly varying fields, as the GEM (for gravitoelectricmagnetic)
approximation or limit.
\setcounter{equation}{8}

Equations (2.6) are of course almost identical to the equations of
classical electromagnetism (EM) for the Coulomb potential $\phi_c$ and
the vector potential $A^j$,
\begin{align}
-\partial^2\phi_c=4\pi \rho_q,\mbox{ } -\partial^2A^j=4\pi \rho_q \mbox{v}^j, \mbox{ }\dot{\phi_c}+\nabla \cdot \vec{A}=0.
\end{align}
Here $\rho_q$ is the charge density and the Lorentz gauge is imposed. The only difference between EM and GEM equations is a factor of $-4$ associated with the gravitomagnetic potential. Moreover the analog of the electric field is the Newtonian or gravitoelectric field $\vec{\mbox{g}}$, while the analog of the magnetic
field $\vec{B}=\nabla\times \vec{A}$ is the gravitomagnetic (or ``frame dragging'') field $\vec{\Omega}$, both defined by
\begin{align}
\vec{\mbox{g}}=-\nabla\phi,\mbox{ } \vec{\Omega}=\nabla\times\vec{h}.
\end{align}

For many systems of interest the GEM equations are easily solved in the
same way as those of EM, for example by the use of a Biot-Savart type law. For a stationary spinning sphere like the earth or (approximately) a neutron star the exterior solutions are \cite{2.3}\cite{2.4}\cite{1.10}
\begin{align}
&\phi=-\frac{GM}{r},\mbox{\space\space\space\space\space\space\space\space\space}\vec{\mbox{g}}=-\frac{Gm}{r^2}\hat{r}\notag\\
&\vec{h}=\frac{2GI}{r^2}\vec{\omega}\times\hat{r},\mbox{\space\space\space\space }\vec{\Omega}=\frac{2GI}{r^3}[3(\hat{r}\cdot\vec{\omega})\hat{r}-\vec{\omega}],
\end{align}
where $\vec{\omega}$ is the spin and $I$ is the moment of inertia of the sphere; for a uniform density sphere $I=(3/5)Mr_s^2$ \cite{1.10}. The fields in Eq. (2.11) are of course time independent.

Another body of theoretical interest is a hollow spherical shell, since the gravitoelectric field vanishes in its interior, while the gravitomagnetic field near the center is about \cite{2.4}
\begin{align}
\vec{\Omega}=\frac{2GM}{r_{hs}}\vec{\omega}.
\end{align}
A gyroscope at the center of such a shell would precess at $\vec{\Omega}/2$ \cite{1.10}.

\section{\label{sec:level1}Schroedinger equation form (SEF) and electromagnetic interaction}
\setcounter{equation}{0}

First we briefly discuss the free scalar Klein-Gordon equation and recast it into a Schroedinger equation form (SEF), which is useful in obtaining the non-relativistic limit \cite{1.9}\cite{1.11}. The SEF is exact and equivalent to the usual Klein-Gordon equation. Surprisingly, it does not appear to be well-known and we have found no reference to it in the literature \cite{1.11}\cite{3.1}.

The free Klein-Gordon Lagrangian and equation are
\begin{align}
L=\varphi^*_{|\mu}\varphi_{|\mu}\eta^{\mu\nu}-m^2\varphi^*\varphi,\mbox{ }\mbox{ }\ddot\varphi-\nabla^2\varphi+m^2\varphi=0.
\end{align}
The time variation of the scalar field $\varphi$ due to
the rest energy $m$ can be separated out by the transformation
\begin{align}
\varphi=e^{-imt}\psi.
\end{align}
Substitution of this into Eq. (3.1) leads to an equivalent form for the Klein-Gordon equation,
\begin{align}
i\dot{\psi}=-\frac{\nabla^2\psi}{2m}+\frac{\ddot{\psi}}{2m}.
\end{align}
Equation (3.3) is exact and has the form of a non-relativistic Schroedinger equation, but contains an extra term, a second derivative, hence our appellation of SEF. An exact solution to Eq. (3.3) is the plane wave $exp(-iE_nt+i\vec{p}\cdot\vec{x})$, where $E_n=E-m$ is the non-relativistic energy.

In a low velocity system the time variation of $\psi$ is associated with non-relativistic kinetic energy,which is much smaller than that of $e^{-imt}$ associated with rest energy. Thus the last term in Eq. (3.3) will be small for positive energy solutions, and can be viewed as a relativistic correction; for negative energy solutions this is obviously not so. To lowest order approximation the relativistic correction corresponds to the standard $\vec{p}^4$ term in the expansion of the relativistic energy.

Note that if the second time derivative in Eq. (3.3) is treated as a small perturbation the Cauchy initial value structure is that of a first order Schroedinger equation. Of course this presents no contradiction with the Cauchy structure of the second order Klein-Gordon equation - so long as we limit ourselves to positive energy solutions \cite{3.2}.

To include electromagnetic interactions we use the standard gauge invariant minimal coupling recipe, $i\partial_\mu\rightarrow i\partial_\mu-eA_\mu$\cite{1.9}\cite{3.3}. The Lagrangian then becomes
\begin{align}
L=(-i\varphi^{*}_{|\mu}-eA_\mu\varphi^*)(i\varphi_{|\nu}-eA_\nu\varphi)\eta^{\mu\nu}-m^2\varphi^*\varphi,
\end{align}
which gives the Klein-Gordon equation in manifestly gauge invariant form,
\begin{align}
(i\partial_\mu-eA_\mu)(i\partial_\nu-eA_\nu)\eta^{\mu\nu}\varphi-m^2\varphi=0.
\end{align}
Equation (3.5) may also be written in a form displaying the electromagnetic coupling separately on the right side,
\begin{align}
\varphi_{|\mu|\nu}\eta^{\mu\nu}+m^2\varphi=&-2ie\varphi_{|\mu}A_{\nu}\eta^{\mu\nu}-ieA_{\mu|\nu}\eta^{\mu\nu}\varphi\notag\\
&+e^2A_\mu A_{\nu}\eta^{\mu\nu}\varphi.
\end{align}
Notice that we use only lower indices on the potential $A_\nu$ and only upper indices on the
Lorentz metric $\eta^{\mu\nu}$, which will prove to be convenient. Also for convenience we use the
Lorentz gauge, without loss of generality, since the Lagrangian is gauge invariant. Thus henceforth we take $eA_{\mu|\nu}\eta^{\mu\nu}=\dot{V}+ e\nabla\cdot\vec{A}=0$, so the second term on the right of Eq. (3.6)
is zero. The Lorentz gauge is particularly appropriate for problems involving a time
independent potential or an external radiation field.

To recast Eq. (3.6) into SEF we make the substitution Eq. (3.2), as before, to obtain
\begin{align}
i\dot{\psi}=&\frac{\vec{\Pi}^2\psi}{2m}+V\psi-\frac{(i\partial_t-V)^2\psi}{2m}, \dot{V}+e\nabla\cdot\vec{A}=0\\
V&\equiv eA^0,\mbox{ }\mbox{ } \vec{p}\equiv i\nabla,\mbox{ }\mbox{ } \vec{\Pi}\equiv\vec{p}-e\vec{A}\notag.
\end{align}

Equivalently we could apply minimal coupling to the free SEF Eq. (3.3) to obtain Eq. (3.7). Eq. (3.7) is an SEF for the spin zero particle interacting with the electromagnetic field in the standard way, with a second derivative relativistic correction. It is exact and useful in obtaining the non-relativistic limit, but apparently does not appear in the literature\cite{3.1}. Note that in Eq. (3.7) there occurs a term $ie\nabla \vec{A}\psi$ in the kinetic energy and a term $i\dot{V}\psi$ in the relativistic correction. In the Lorentz gauge that we are using they cancel and are not really present in the SEF of Eq. (3.7).

Eq. (3.7) contains a somewhat subtle feature related to conservation of probability and reality of the energy, which we next consider. In the Schrodinger equation of nonrelativistic quantum mechanics,
\begin{align}
i\dot{\psi}=H\psi=\frac{p^2}{2m}\psi+V\psi,
\end{align}
the potential $V$ is usually taken to be real, thus making the Hamiltonian Hermitian; this guarantees that its energy eigenvalues are real and that probability is conserved. An equivalent way to analyze probability conservation is in terms of a probability density and associated 3-vector current. These are defined as
\begin{align}
\rho=\psi^*\psi,\space\vec{j}=\frac{-1}{2m}\psi^*\overleftrightarrow{\nabla}\psi\equiv\frac{-i}{2m}\left(\psi^*\nabla\psi-\nabla\psi^*\psi\right).
\end{align}
From Eq. (3.8), we readily obtain the conservation equation
\begin{align}
\dot{\rho}+\nabla\cdot\vec{j}=i\psi^*(V-V^*)\psi=0.
\end{align}
Thus probability is conserved if the potential is real.

For the relativistic Klein-Gordon equation interacting with the electromagnetic field the analysis of probability conservation proceeds in a similar way. The 4-vector current is defined as
\begin{align}
j^\mu&=\frac{i}{2m}\left(\varphi^*\varphi^{|\mu}-\varphi^{*|\mu}\varphi\right) -\frac{eA^\mu}{m}\varphi^*\varphi\notag\\
&\equiv\frac{i}{2m}\varphi^*\overleftrightarrow{\partial}^\mu\varphi-\frac{eA^\mu}{m}\varphi^*\varphi.
\end{align}
Conservation follows from the Klein-Gordon Eq. (3.5) or (3.6), and is expressed as $j^\mu_{\mbox{\space}|\mu}=0$. The same $j^\mu$, up to a constant factor, follows from the Lagrangian Eq. (3.4) as the electromagnetic current. We normalize the current so that the 3-vector part is consistent with the Schrodinger current. In terms of $\psi$ defined in Eq. (3.2) the density and 3-vector current are
\begin{align}
j^0&=\left(1-\frac{V}{m}\right)\psi^*\psi+\frac{i}{2m}\psi^*\overleftrightarrow{\partial}^0\psi,\notag\\ j^k&=\frac{i}{2m}\psi^*\overleftrightarrow{\partial}^\mu\psi-\frac{eA^k}{m}\psi^*\psi.
\end{align}
Thus the conserved probability density $j^0$ is not the same as the Schrodinger density $\psi^*\psi$ in Eq. (3.9), although it is simply related to it, as we now discuss.

It is instructive to consider the case of an energy eigenfuntion $\psi(\vec{x},t)\rightarrow e^{-iEt}\psi(\vec{x})$. Then the probability density in Eq. (3.12) becomes, approximately,
\begin{align}
j^0=\left(1+\frac{E-V}{m}\right)\psi^*(\vec{x})\psi(\vec{x}) \cong\left(1+\frac{\vec{p}^2}{2m^2}\right)\psi^*(\vec{x})\psi(\vec{x}).
\end{align}
We can view Eq. (3.13) as a relativistic renormalization of the wave function since the factor $(1+p^2/2m^2)$ is a Lorentz volume contraction factor. This is easy to see if we write it in terms of the velocity, defined as $\vec{\mbox{v}}=\vec{p}/m$,
\begin{align}
1+\frac{\vec{p}^2}{2m^2}=1+\frac{\vec{\mbox{v}}^2}{2}\cong\frac{1}{\sqrt{1-\vec{\mbox{v}}^2}}.
\end{align}

We will illustrate the importance of the wave function renormalization for the important example of a slowly moving particle in a time independent potential $eA^0 = V$ and zero 3-vector potential \textendash\space that is $\dot{V}= 0$ and $\vec{A}=0$. This calculation is the spinless analog of the well-known hydrogen fine structure calculation based on a non-relativistic reduction of the Dirac equation\cite{shankar}. To seek energy eigenvalues we again set $\psi(\vec{x},t)\rightarrow e^{-iEt}\psi(\vec{x})$ and Eq. (3.7) becomes
\begin{align}
(E-V)\psi=\frac{p^2}{2m}\psi-\frac{1}{2m}(E-V)^2\psi.
\end{align}
The operators on the left side and the first term on the right side are of order $O(m\mbox{v}^2)$ and the last term on the right is $O(m\mbox{v}^4)$; we will work to only this order. We iterate Eq. (3.15) and obtain
\begin{align}
E\psi=&\frac{p^2}{2m}\psi+V\psi-\frac{1}{2m}(E-V)\frac{p^2}{2m}\psi\notag\\
     =&\frac{p^2}{2m}\psi+V\psi-\frac{1}{2m}\left[(E-V),p^2/2m\right]\psi\notag\\
     &-\frac{p^2}{4m^2}(E-V)\psi\notag\\
     =&\frac{p^2}{2m}\psi+V\psi-\frac{p^4}{8m^3}\psi+\frac{1}{2m}\left[V,p^2/2m\right]\psi.
\end{align}
However the last term in Eq. (3.16) is not Hermitian, as is easily verified, so it is not a consistent eigenvalue equation for the real energy. This is because the wave function $\psi$ is not correctly normalized. We renormalize it according to Eq. (3.13) by defining a wave function that is normalized like the Schrodinger wave function,
\begin{align}
\psi_S\cong(1+p^2/4m^2)\psi,\mbox{\space\space}\psi\cong(1-p^2/4m^2)\psi_S.
\end{align}
Substituting Eq. (3.17) into Eq. (3.16) we obtain for $\psi_S$, correct to order $O(m\mbox{v}^4)$,
\begin{align}
E\psi_S=&(1+p^2/4m^2)\left(\frac{p^2}{2m}+V\right)(1-p^2/4m^2)\psi_S\notag\\
&-\frac{p^4}{8m^3}\psi_S+\frac{1}{2m}\left[V,p^2/2m\right]\psi_S\notag\\
=&\left(\frac{p^2}{2m}+V\right)\psi_S+\left[p^2/4m^2,V\right]\psi_S-\frac{p^4}{8m^3}\psi_S\notag\\
&+\frac{1}{2m}\left[V,p^2/2m\right]\psi_S\notag\\
=&\left(\frac{p^2}{2m}+V\right)\psi_S-\frac{p^4}{8m^3}\psi_S.
\end{align}
Thus the non-Hermitian term has been cancelled by the wave function renormalization, so the resulting equation will have real energy eigenvalues. It is worth noting that the procedure leading to Eq. (3.18) is analogous to that leading to the hydrogen fine structure energy, but the result is simpler since there are no spin-orbit and Darwin terms\cite{shankar}.

In summary, Eq. (3.7) is gauge invariant, and contains only Hermitian operators when the gauge is carefully chosen and wave function renormalization is applied. Similar considerations will be used in sections 4 to 6.

\section{\label{sec:level1}Gravitational interaction}
\setcounter{equation}{0}
For a scalar field the interaction with gravity is obtained by
making the Klein-Gordon equation generally covariant.
To do this we need only multiply the flat space Lagrangian by the square
root of the absolute value of the metric determinant
(denoted $\sqrt{\mbox{g}}$) to form a scalar density, replace the
Lorentz metric $\eta^{\mu\nu}$ by the Riemannian metric $\mbox{g}^{\mu\nu}$,
and substitute covariant derivatives for ordinary derivatives
\cite{4.1}. Then, in covariant form, the Lagrangian in Eq. (3.4) becomes a density,
\begin{align}
L=\sqrt{\mbox{g}}[(-i\varphi^*_{|\mu}-eA_\mu\varphi^*)(i\varphi_{|v}-eA_\mu\varphi)\mbox{g}^{\mu\nu}-m^2\varphi^*\varphi],
\end{align}
and the Klein-Gordon Eq. (3.6) becomes,
\begin{align}
\varphi_{|\mu\parallel\nu}\mbox{g}^{\mu\nu}+m^2\varphi=&-2ie\varphi_{|\mu}A_{\nu}\mbox{g}^{\mu\nu}-ieA_{\mu\parallel\nu}\mbox{g}^{\mu\nu}\varphi\notag\\
&+e^2A_\mu A_\nu\mbox{g}^{\mu\nu}\varphi.
\end{align}
The double slash here denotes a covariant derivative and the single slash an ordinary
derivative; for a scalar they are the same. The covariant d¡¦Alembertian and divergence
terms are conveniently expressed as \cite{2.1},
\begin{align}
\varphi_{|\mu\parallel\nu}\mbox{g}^{\mu\nu}=\varphi_{\parallel\alpha}^{\mbox{\space\space}\parallel\alpha}=\frac{1}{\sqrt{\mbox{g}}}\left(\sqrt{\mbox{g}}\varphi_{|\mu}\mbox{g}^{\mu\nu}\right)_{|\nu},\notag\\ A_{\mu\parallel\nu}\mbox{g}^{\mu\nu}=A^{\nu}_{\mbox{\space}\parallel\nu}=\frac{1}{\sqrt{\mbox{g}}}(\sqrt{\mbox{g}}A^\nu)_{|\nu}.
\end{align}

We have here assumed only the simplest coupling to gravity, consistent with the equivalence principle \cite{2.4}. Additional terms involving the Riemann tensor are also possible since they vanish in flat space and are thus consistent with the flat space Klein-Gordon equation \cite{1.7}.

Covariant electromagnetism is gauge invariant so we again choose for convenience the Lorentz gauge, $A_{\parallel \alpha}^{\mbox{\space\space}\parallel\alpha}=0$; the second term on the right of Eq. (4.2) is then zero. Recall that we also using the Lorentz gauge for the gravitational field, Eq. (2.4).

We next substitute Eq. (4.3) into Eq. (4.2), expand the metric to first order, $\mbox{g}^{\mu\nu}=\eta^{\mu\nu}-h^{\mu\nu}$, and apply the Lorentz gauge condition Eq. (2.4); this yields
\begin{align}
\varphi&_{|\mu|\nu}\eta^{\mu\nu}+m^2\varphi\notag\\
=&\varphi_{|\mu|\nu}h^{\mu\nu}+(-2ie\varphi_{|\mu}A_\nu+e^2A_\mu A_\nu\varphi)(\eta^{\mu\nu}-h^{\mu\nu}).
\end{align}
This surprisingly simple result Eq. (4.4) only holds in the Lorentz gauge for both the electromagnetic and gravitational fields. Eq. (4.4) holds to all orders in the electromagnetic field, and to first order in the gravitational field. The right side includes the electromagnetic terms of (3.6), an additional term for coupling to the gravitational field, and two ¡§mixed¡¨ terms for coupling to the product of the electromagnetic and gravitational fields. The mixed terms are novel and may be of possible interest in connection with very strong electromagnetic fields, such as those found in laboratory magnets or intense laser light, as we will discuss in section 6.

To recast Eq. (4.4) in SEF we separate out the rest energy time dependence using Eq. (3.2) as before. Straightforward manipulation leads to a generalization of Eq. (3.7) to include gravitational interaction terms, which we write as
\begin{align}
i\dot{\psi}=&\frac{\vec{\Pi}^2\psi}{2m}+V\psi-\frac{(i\partial_t-V)^2\psi}{2m}+H_{\text{g}}\psi+H_{\text{g}e}\psi.
\end{align}
The term $H_{\text{g}}\psi$ contains interactions with only the gravitational field, and in the GEM approximation of section 2 it is
\begin{align}
H_{\text{g}}\psi=\phi\left(m\psi-\frac{\nabla^2\psi}{m}+2(i\dot{\psi})-\frac{\ddot{\psi}}{m}\right)\notag\\
+(\vec{h}\cdot\vec{p})\left(\psi+\frac{(i\dot{\psi})}{m}\right).
\end{align}
The mixed term $H_{\text{g}e}\psi$ contains interactions with the product of the gravitational and electromagnetic fields, and in the same approximation it is
\begin{align}
H_{\text{ge}}\psi=&\Big(-2\phi V(\psi-\frac{i\dot{\psi}}{m})-2\phi\frac{e\vec{A}\cdot(\vec{p}\psi)}{m}\notag\\
&-e\vec{A}\cdot\vec{h}(\psi-\frac{(i\dot{\psi})}{m})-\frac{V\vec{h}\cdot(\vec{p}\psi)}{m}\Big)\notag\\
&+\left(\phi\frac{V^2\psi}{m}+\phi\frac{e^2\vec{A}^2\psi}{m}+\frac{Ve(\vec{A}\cdot\vec{h})\psi}{m}\right).
\end{align}
Eq.s (4.5) to (4.7) are a restatement of Eq. (4.4) in terms of the wave function $\psi$ and in the GEM approximation for the gravitational field. The only approximations made so far are the GEM of section 2 for the external gravitational field, which assumes a slow moving source. The electromagnetic field is treated exactly and there is no low velocity approximation for the particle.

\section{\label{sec:level1}Non-relativistic limit}
\setcounter{equation}{0}

In this section we consider the low velocity limit of Eq.s (4.5) to (4.7) to obtain our main, remarkably simple, result. The four terms in the first parenthesis of Eq. (4.6) are $\phi$ times, respectively, $O(m)$, $O(m\mbox{v}^2)$, $O(m\mbox{v}^2)$, and $O(m\mbox{v}^4)$; we will work to $O(m\mbox{v}^2)$ and drop the last term. In the second parenthesis the terms are h times, respectively, $O(m\mbox{v})$ and $O(m\mbox{v}^3)$; again we drop the last term. We then do the same order counting in Eq. (4.7) and combine the remaining terms with those of Eq. (4.6) to obtain an SEF,
\begin{align}
i\dot{\psi}=&\left[\frac{\vec{\Pi}^2\psi}{2m}+V\psi-\frac{(i\partial_t-V)^2\psi}{2m}\right]\notag\\
&+\phi m\left(1+\frac{2\vec{\Pi}^2}{m^2}\right)\psi+\vec{h}\cdot\vec{\Pi}\psi.
\end{align}
Thus the Newtonian field $\phi$ couples to the particle mass with a small correction factor, and the gravitomagnetic field $\vec{h}$ couples to the kinematic momentun $\vec{\Pi}$. We will discuss
the gravitomagnetic coupling at length in the next section.

The factor preceding the Newtonian field in Eq. (5.1) is interesting. It may be written in terms of the particle velocity as $(1+ 2\mbox{v}^2)$, with the velocity defined as $\vec{\mbox{v}}=\vec{\Pi}/m$. We can show that $(1+ 2\mbox{v}^2)\phi$ is simply the potential that the moving particle sees in its rest frame. To see this we apply a Lorentz transformation in the x direction to the metric perturbation,
\begin{align}
&h'^{\mu\nu}=\alpha^\mu_{\mbox{\space}\alpha} \alpha^\nu_{\mbox{\space}\beta}h^{\alpha\beta},\mbox{\space\space} \alpha^\mu_{\mbox{\space}\alpha}=
\begin{pmatrix}
  \gamma & -\mbox{v}\gamma  \\
   -\mbox{v}\gamma     & \gamma
\end{pmatrix},\notag\\
&\gamma=\frac{1}{\sqrt{1-\mbox{v}^2}}.
\end{align}
Ignoring the small off-diagonal terms of the metric we see from Eq. (2.5) that $h^{00}$ and
thus the Newtonian potential transform like
\begin{align}
\phi'=(a^0_{\mbox{\space}0}a^0_{\mbox{\space}0})\phi+(a^0_{\mbox{\space}1}a^0_{\mbox{\space}1})\phi=\gamma^2(1+\mbox{v}^2)\phi\cong(1+2\mbox{v}^2)\phi.
\end{align}

Finally, we emphasize that Eq. (5.1) contains only Hermitian terms in the gravitational interactions; recall also that in section 3 we showed that the rest of the equation contains only manifestly Hermitian terms if the Lorentz gauge is used and the wave function is properly renormalized; thus no problem arises with conservation of probability and reality of the energy eigenvalues.

\section{\label{sec:level1}Gravitomagnetic physical effects}
\setcounter{equation}{0}

We will briefly discuss some physical effects implied by Eq. (5.1), concentrating on gravitomagnetism. Our purpose is only to provide some simple parametrizations and sample numerical estimates. A specific and detailed discussion of any laboratory experiments or astrophysical observations is beyond our present scope, so we leave it as a challenge to experimentalists to design and analyze specific experiments.

We will consider here only neutral systems, perhaps containing charged particles, since electromagnetic effects on a system with net charge generally swamp gravitational forces by many orders of magnitude \cite{1.1}.

The gravitoelectric or Newtonian term $\phi m\psi$ in Eq. (5.1) is generally much the dominant gravitational interaction. Its effects have been seen in diverse experiments, notably on neutrons in the earth's gravitational field, which make it clear that nucleons do indeed fall and behave as expected \cite{1.75}\cite{1.76}! That term is also responsible for most of the phase shift seen in recent atomic beam interferometry gravity experiments \cite{1.8}. Such experiments may be able to test the equivalence principle to $10^{-15}$ or better; however the phase shifts, including small subtle corrections, may be calculated more easily and
accurately with the semi-classical approach of Dimopoulous et. al. that treats the atoms as neutral particles, rather than the approach we adopt here \cite{1.8}.

Our present interest is more on the gravitomagnetic term $\vec{h}\cdot\vec{\Pi}\psi$ of Eq. (5.1). The first part, $\vec{h}\cdot\vec{p}\psi$, is the analog of the EM coupling to the vector potential, as we see by writing out the kinetic energy term \cite{6.1},
\begin{align}
\frac{\vec{\Pi}^2\psi}{2m}=\frac{\vec{p}^2\psi}{2m}-\frac{e\vec{A}\cdot\vec{p}\psi}{m}+\frac{ie\nabla\cdot\vec{A}\psi}{2m}+\frac{e^2\vec{A}^2\psi}{2m}.
\end{align}
From the second term of Eq. (6.1) the correspondence between magnetic and gravitomagnetic couplings is thus evident,
\begin{align}
\vec{h}\Leftrightarrow-\frac{e}{m}\vec{A}.
\end{align}

It is illuminating to take the quantum system to be a fictitious atom containing a scalar electron of mass $m_e$, which we then place in
$\vec{B}$ and/or $\vec{\Omega}$ fields that are approximately constant over the size of the atom. In this case we may conveniently choose the potentials as
\begin{align}
\vec{A}=\frac{1}{2}\vec{B}\times\vec{r},\mbox{ } \vec{h}=\frac{1}{2}\vec{\Omega}\times\vec{r},
\end{align}
where $\vec{r}=0$ is the center of mass of the atom. The magnetic
and gravitomagnetic interactions are then, from Eq.s (5.1) and (6.1),
\begin{subequations}
\begin{align}
-\frac{e}{m_e}\vec{A}\cdot\vec{p}\psi=-\frac{e}{2m_e}\vec{B}\times\vec{r}\cdot\vec{p}\psi =-\frac{e}{2m_e}\vec{B}\cdot\vec{r}\times\vec{p}\psi\notag\\
=-\frac{e}{2m_e}\vec{B}\cdot\vec{L}\psi, \\
\vec{h} \cdot \vec{p} \psi = \frac{1}{2} \vec{\Omega} \times\vec{r}  \cdot \vec{p} \psi = \frac{1}{2}\vec{\Omega} \cdot \vec{r} \times \vec{p} \psi = \frac{1}{2} \vec{\Omega} \cdot \vec{L} \psi.
\end{align}
\end{subequations}
Here the orbital angular momentum $\vec{L}$ is taken with respect to the center of mass. Thus Eq. (6.4a) means that orbital angular momentum produces a magnetic moment $\vec{\mu}=(e/2m_e)\vec{L}$, since Eq. (6.4a) is the energy of a magnetic moment in a magnetic field. In the same manner we interpret Eq. (6.4b) to say that orbital angular momentum produces a gravitomagnetic moment, $\vec{\mu}_{grav}=(-1/2)\vec{L}$. It is clear from the correspondence that, since a magnetic moment precesses at the Larmor frequency $eB/2m_e$ in a $B$ field, the gravitomagnetic moment will precess in a gravitomagnetic field with frequency $\Omega/2$, and in the opposite direction.

The classical analog of the gravitomagnetic coupling in (6.4b) leads to the Lense-Thirring precession of a gyroscope in the field of the earth as observed in the Gravity Probe B experiment \cite{1.6}\cite{1.10}. Thus a quantum system such as our fictitious atom should undergo the same precession as a classical system in a gravitomagnetic field. This should be expected for a fundamentally geometric effect, according to general relativity.

We note that a similar analogy holds for classical orbital motion: a charged particle in a constant $B$ field (and no $E$ field) orbits at the cyclotron frequency $eB/m_e$, whereas a massive particle in a constant $\Omega$ field (and no g field) orbits at frequency $\Omega$ \cite{2.4}. It is of course easy to produce a region of $\vec{B}\cong constant$ $\vec{E}=0$ in the lab; it is similarly easy to produce a region of $\vec{\Omega}\cong constant$ $\vec{\text{g}}=0$ by using the interior of a massive spherical shell, as we noted in section 2 \cite{2.4}. Thus a particle near the center of such a hollow shell will orbit at $\Omega=(2GM/r_{hs})\omega$.

It is of some interest to estimate the gravitomagnetic precession frequency of the fictitious atom near the surface of two bodies of interest, the earth and a neutron star, both approximated by a spinning sphere with a gravitomagnetic field given by Eq. (2.11). In
order of magnitude
\begin{align}
\Omega \approx \frac{2G \left( M r^2_b\right)}{r^3} \omega = \frac{r_s r^2_b}{r^3}\omega,
\end{align}
where $r_s=2GM$ is the body's Schwarzschild radius and $r_b$ is its actual radius. For the earth, $r\approx r_b=6\times10^3$km, $r_s=1.8$cm, $\omega=7.3\times10^{-5}$rad/s. For a typical rapidly
spinning neutron star we take $r\approx r_b\approx10$km, $r_s\approx1$km,
$\omega=10^2$rad/s. Then the precession frequencies and associated energies are of order
\begin{subequations}
\begin{align}
\Omega \approx 10^{-13} \text{rad/s,\space}E_\text{char} \approx \hbar \Omega \approx 10^{-28} \mbox{eV} \mbox{ (earth )}\\
\Omega \approx 10 \text{rad/s,\space}E_\text{char} \approx \hbar \Omega \approx 10^{-14} \mbox{eV} \mbox{ (neutron star)}
\end{align}
\end{subequations}
The energies in Eqs. (6.6) are of course extremely small compared to typical atomic energies, even hyperfine energies; in any situation where GEM effects directly compete with EM effects they will naturally be swamped by many orders of magnitude.

The other part of the gravitomagnetic term in Eq. (5.1), $-\vec{h}\cdot e\vec{A}\psi$, might conceivably be of interest for some terrestrial or astrophysical environments if the vector potential $\vec{A}$ is sufficiently large. Two example systems of interest, obtainable in a
terrestrial laboratory, are a high field magnet and an intense laser beam \cite{6.2}\cite{6.3}. We first consider a quantum system such as our fictitious atom in a large $\vec{B}$ field. The perturbation energy,
with the use of Eq. (2.10), is of order
\begin{align}
\Delta E&\approx \langle\vec{h}\cdot e\vec{A}\rangle\approx\Omega e Br^2_q\notag\\
&=\Omega\left(\frac{eB}{m_e}\right)r^2_qm=\left(\frac{eB}{m_e}\right)\left(\frac{r_q}{\lambda_c}\right)\left(\frac{r_q}{c}\right)\Omega. \end{align}
Here $r_q$ is the characteristic size of the quantum system, $\lambda_c=\hbar/m_ec\approx2.43\times10^{-12}$m is the Compton wavelength of the electron, and we have included a factor of c to make the dimensions clear. For a numerical estimate we take $r_q\approx10^{-9}$m, $\Omega\approx10^{-13}$rad/s from Eq. (6.6a), $B\approx10^2$T, and express the Bohr magneton as $e/2m_e=(5.8\times10^{-5}$eV/T$)$. Then the
energy and associated frequency are of order,
\begin{align}
\Delta E\approx10^{-32}\mbox{eV}, \Delta\omega\approx10^{-17}\text{rad/s\space\space(atom in terrestrial B)}
\end{align}
This is even smaller than the values in Eq. (6.6a) so the presence of the magnetic field would not appear to produce a larger effect of interest.

However if we suppose a much larger quantum system, perhaps Cooper pairs
 in a Josephson junction of nearly macroscopic size $r_q\approx10^{-6}$m,
 then the above values increase by six orders of magnitude and possibly
 become more interesting \cite{6.4},
\begin{align}
\Delta E\approx10^{-26}\mbox{eV}, &\Delta\omega\approx10^{-11}\text{rad/s}\notag\\
 &\text{\space\space(large quantum system in terrestrial B)}
\end{align}
We leave it to experimentalists to consider such larger systems.

Finally we consider a fictitious atom in an intense laser beam
of frequency $\omega_l$ and electric field $E$. For a radiation field
$A\approx E/\omega_l$ and we may estimate the perturbation as before
\begin{align}
\Delta E&\approx \langle\vec{h}\cdot e\vec{A}\rangle\approx\Omega
\left(\frac{eE_l}{\omega_l}\right)r_q\notag\\
&=\left(\frac{eE_l}{m_e\omega_l}\right)\left(\frac{r_q}{\lambda_c}\right)\Omega=a_0\left(\frac{r_q}{\lambda_c}\right)\Omega.
 \end{align}
The quantity $eE_l/\omega_l$ is the approximate energy transfer to the electron in one cycle of the laser beam, so the dimensionless parameter
$a_0\equiv eE/m_e\omega_l$ is the fraction of the electron rest energy absorbed from the laser beam in one cycle; it is a measure of how relativistic the quantum system can become, and is thus also a convenient laser intensity parameter \cite{6.3}. Note that we may also express
$a_0$ in terms of the Schwinger critical field, at which spontaneous pair production is relevant, as $a_0=(E/E_{SC})(m_e/\omega_l)$ \cite{6.5}.

If we take as rough examples $a_0=1$ and $r_q\approx10^{-9}$m we obtain from Eq. (6.10)
\begin{align}
\Delta\omega\approx10^{-11}\text{rad/s\space}&, \Delta E\approx10^{-26}\mbox{eV}\notag\\ &\text{\space\space(atom in terrestrial laser)}
\end{align}
Of course our approximations do not hold for $a_0$ much larger than 1, and are probably not even very accurate for $a_0$ of order 1. The values in Eq. (6.11) are larger than those in Eq. (6.6a) so the presence of the laser field may increase the gravitomagnetic effect and possibly be of interest to experimentalists.

For a larger quantum system, for example a Josephson junction as
mentioned above, the values might increase by perhaps three orders of
magnitude, according to Eq. (6.10), to about
\begin{align}
\Delta\omega\approx10^{-8}\text{rad/s\space}&, \Delta E\approx10^{-23}\mbox{eV}\notag\\ &\text{\space\space(larger system in terrestrial laser)}
\end{align}
We note incidentally that measurements of the Josephson constant $2e/h$ (relating voltage to frequency) reach a precision of about $2\times10^{-11}$ \cite{6.7}.

There is a major difficulty associated with detecting gravitomagnetic effects that we have not yet mentioned; any rotation of the laboratory apparatus will in general compete with and swamp gravitomagnetic effects via the Sagnac effect \cite{6.8}. Thus any experiment would need to provide a way to suppress or separate such rotational effects at
about the level of $10^{-13}$rad/s according to Eq. (6.6a).

\section{Summary and Conclusions}
Our study of the interaction of a spinless quantum particle with an
electromagnetic field and a weak gravitational field has lead to a rather simple non-relativistic limit expression that is useful for considering possibly observable gravitomagnetic effects. The laboratory observation of such effects is clearly a very difficult prospect, as our illustrative examples indicate, but would be of fundamental interest.

\section*{Acknowledgements}
This work was partially supported by NASA grant 8-39225 to Gravity Probe B and by NSC of Taiwan under Project No. NSC 97-2112-M-002-026-MY3 and by US Department of Energy under Contract No. DE-AC03-76SF00515. Thanks go to Robert Wagoner, Francis Everitt, and Alex Silbergleit and members of the Gravity Probe B theory group for useful discussions, and to Mark Kasevich of the Stanford physics department for interesting comments on atomic beam interferometry and equivalence principle experiments. Pisin Chen also thanks Taiwan's National Center for Theoretical Sciences for the support. Kung-Yi Su provided generous checking and technical help with
the manuscript.


\end{document}